\documentclass[aps,pra,onecolumn,floatfix]{revtex4}
\usepackage{graphicx}
\usepackage{amsmath}
\usepackage{lscape}
\newcommand{\op}[1]{\hat{#1}}
\newcommand{\grad}{\nabla}
\renewcommand{\-}{\nonumber\\}
\newcommand{\expec}[1]{\left\langle #1 \right\rangle}

\newcommand{\authornote}[1]{}

\newcommand{\comment}[1]{}
\renewcommand{\Im}{\mbox{Im}}
\renewcommand{\Re}{\mbox{Re}}
\newcommand{\sub}[1]{_{\mbox{\tiny #1}}}

\begin{document}
\title{Quantum dynamics of electronic excitations in biomolecular chromophores: role of the protein environment and solvent}
\author{Joel Gilmore and Ross H. McKenzie
\\ {\em \small Department of Physics, University of Queensland, Brisbane 4072 Australia}
}
\date{\today}

\begin{abstract}
We consider continuum dielectric models as  minimal models
to understand the effect of a surrounding globular protein and
solvent on the quantum dynamics of electronic excitations in a
biological chromophore. 
For these models we derive   expressions for the frequency dependent
spectral density which describes the coupling of the electronic
levels in the chromophore to its environment.
The magnitude and frequency dependence of the
spectral density determines whether or not the quantum
dynamics is coherent or incoherent, and thus whether on
not one can observe quantum interference effects such as 
Rabi oscillations.
We find the contributions to the spectral density
from each component of the chromophore environment: the
bulk solvent, protein, and water bound to the protein.
The relative importance of each component
to the quantum dynamics of the chromophore
is determined by the time scale on which
one is considering the dynamics.
Our results provide a natural explanation and model for
the different time scales observed in the spectral density
extracted from the solvation dynamics
probed by ultra-fast laser spectroscopy techniques such
as the dynamic Stokes shift and 
three pulse photon echo spectroscopy.
Our results are used to define under what conditions the
dynamics of the excited chromophore is dominated by the surrounding protein
and when it is dominated by dielectric fluctuations in the 
solvent. We show that even when the chromophore is shielded
from the solvent by the protein ultra-fast solvation
can be dominated by the solvent. Hence, we suggest that the
ultra-fast solvation recently seen in some biological chromophores
should not necessarily be assigned to ultra-fast protein dynamics.
The magnitudes of the spectral density that we estimate from our
continuum models and extracted from experiment suggest that
most quantum dynamics of electronic excitations is incoherent.
A possible exception is transfer of excitons between neigbouring 
chromophores in photosynthetic systems.

\end{abstract}


\maketitle

\section{Introduction}

The functionality of many proteins is associated with a small subsystem or active site such as a heme group, a couple of amino acids involved in proton transfer, or a co-factor such as an optically active molecule (chromophore).
There are a diverse range of optically active molecules that have an important biological function \cite{Helms02}.  Examples include retinal (involved in vision), green fluorescent protein and porphyrins (photosynthesis).
For these chromophores, the protein acts as a transducer
 which converts optical excitation of the chromophore
 into a change such as an electrical signal or conformational change that
in turn brings about the desired biological function.
 Many of these transducers  operate with speeds, specificities, and efficiencies
which nanotechnologists are striving to mimic \cite{Sarikaya03}.  

The dynamics of a protein involves thousands of degrees of freedom and at room temperature can be described by classical mechanics and modelled
using molecular dynamics methods.
 In contrast, the functional subsystem involves only a few quantum states
and their dynamics must    
 be described quantum mechanically.
This has led to considerable effort at developing hybrid QM/MM
(quantum mechanical-molecular mechanical) methods \cite{Warshel01QRB,Groenhof04JACS}. 
 In most cases the change in quantum state associated with the functional event (e.g., transition to an excited electronic state)
is associated with a change in the electric dipole moment of the subsystem.
Since the protein contains polar residues and is surrounded by a highly polar solvent
 (water)\cite{Gilmore06CPL,Gilmore05JPCM,Edsall83AP,Wand01NSB} there is a strong interaction between the functional subsystem and its environment. Consequently, the environment can have a significant effect on the quantum dynamics
of the subsystem.
Indeed chromophores such as retinal, Photoactive Yellow 
Protein\cite{Groenhof04JACS},  Green Flourescent Protein 
exhibit distinctly different dynamics in solution, in the gas phase,
 and in the protein environment.\cite{Vengris04BJ,Zimmer02}
For example, the speed, efficiency, and selectivity
with which excited retinal undergoes a conformational change
are all significantly less in water than in the protein environment.
\cite{Edsall83AP,Mattos02TBS,Halle04PTRSLB}
This interplay between quantum and classical dynamics raises a number of questions of fundamental
interest. On what length and time scales does the crossover from quantum to classical behaviour occur? When are quantum mechanical effects such as coherence (i.e., superposition states), entanglement, tunneling, or interference necessary for biological function?\cite{Davies04,Engel07N} 
 What aspects and details of the structure and dynamic properties of the protein are crucial to biological function? 

\subsection{Biomolecular chromophores}

\begin{figure}
\begin{center}
  \includegraphics[width=3.5in]{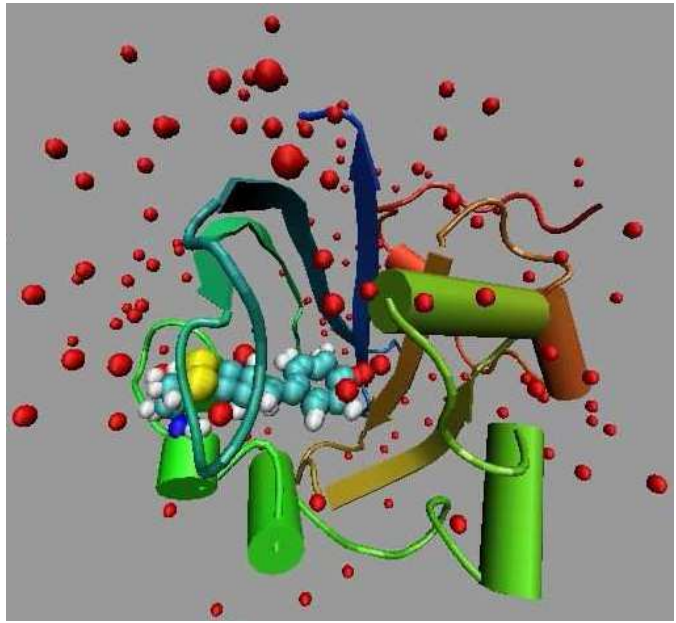}\\
  \caption{The chromophore, protein and bound water in photoactive yellow protein (PYP).  The isolated spheres represent the bound water, the chromophore is shown by its van de Waals
 surface, and the protein by a cartoon representation.
  Observe that the protein surrounding the
chromophore reduces the contact of the chromophore with the surrounding
bulk water.
 Generated from the Protein Database 3PYP.pdb\cite{Genick98N}.}
\label{fig:pyp}
\end{center}
\end{figure}

  Most chromophores are large conjugated organic molecules which are surrounded by a protein which in turn is surrounded by a solvent.
 Figure \ref{fig:pyp} shows the photoactive yellow protein (PYP), including the chromophore and the so-called 
``bound water'' molecules which reside with comparatively long lifetimes on the surface of the protein \cite{Jordanides99JPCB}.
  Most chromophores have large dipole moments which change significantly upon optical excitation, leading to significant relaxation
 of the polarisable environment. 
 Combined chromophore-protein-solvent systems exhibit a broad range of time, length and energy scales (see Figure \ref{fig:timescales}).
Typical values of different time scales are shown
 in Table 1 in the Appendix.  
 
 In this paper, we focus on minimal model Hamiltonians,
since we are seeking to understand general qualitative features,
identify crucial parameters for understanding qualitative
changes in behaviour.
We specifically consider chromophores which can be described as two level systems (TLS), e.g., we only
need consider the ground and first excited state.
 The models proposed here can also be extended to include internal nuclear 
dynamics of the chromophore,
 such as conformational change \cite{Gilmore05JPCM}.

Questions of quantum coherence and the role of the environment are particularly pertinent and controversial in photosynthetic systems
\cite{Chachisvilis97JPCB,Trinkunas01PRL,Brixner05N,Wohlleben}.
 It has sometimes been claimed that the excitons
  within the light harvesting
 rings are quantum mechanically coherent over some or all
of the chromophores (sometimes as many as 32) within the ring.
It has also been suggested that such coherence 
is important for optimum performance
 of the system\cite{Hu97JPCB,Pullerits96JPC,Monshouwer97JPCB}.
 On the other hand,
 inter-ring transfer of excitons is incoherent which ensures the
 desirable feature of irreversible transport of energy towards
 the reaction centre.
Recently, the question of delocalisation of excitons over several base pairs
in DNA has be studied, motivated by a desire to understand UV damage of DNA.\cite{BuchvarovPNAS07}

\begin{figure}
\begin{center}
  \includegraphics[width=15cm]{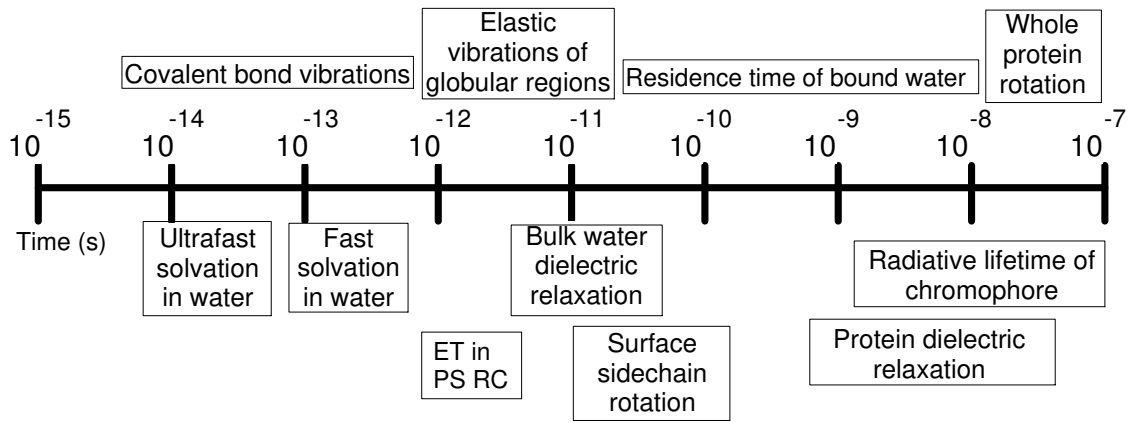}\\
  \caption{Schematic representation of the time scales of various processes in proteins and solutions. ET stands for electron transfer, PS RC for photosynthetic reaction centre.  Specific numbers and references
can be found in the Appendix. }\label{fig:timescales}
\end{center}
\end{figure}

\subsection{Quantum dynamics, decoherence, and the spin-boson model}

\begin{table}[htdp]
\caption{Comparison of the matrix element $\Delta$ which
couples two quantum states
 for various processes in proteins
with the solvation rates due to the interaction
of the quantum system with different parts of
its environment.
The quantum dynamics of the process will be determined
largely by the part of the environment which undergoes
solvation relaxation at a rate comparable to $\Delta$.  LHI and LHII refer to light harvesting complexes I and II in photosynthetic purple bacteria.
}
\begin{center}
\begin{tabular}{|c|c|c|}
\hline
Process & $\Delta$ energy (meV) & Ref.\\
\hline
F\"orster coupling between chromophores in FRET spectroscopy
&  0.2-2 &  \\
Interring F\"orster coupling between chromophores in LHI and LHII
& 0.3 & \cite{Hu97JPCB} \\
Intraring F\"orster coupling between two chlorophyl molecules in LHI
& 50-100 & \cite{Hu97JPCB}\\
F\"orster coupling between infrared amide modes in proteins
& 0.1-1 & \cite{Cho05JCP}\\
Electron transfer in photosynthetic reaction centre (PRC)
&  1-10 & \cite{Zhang98PNAS} \\
Electron transfer in  PRC radical cation
&  10-100 & \cite{Reimers96CP} \\
Electron transfer in proteins & $10^{-4} - 10^{-2}$
 & \cite{Kawatsu02JPCB} \\
Proton transfer
&   0.05 & \cite{Borgis96JPC}  \\
Level crossing for non-radiative decay
&  40 & \cite{Ben-Nun98FD} \\
Solvation rate due to bulk  water &  10 &  \cite{Lang99JCP} \\
Solvation rate due to bound  water &   0.1-1 & \cite{Peon02PNAS} \\
Solvation rate due to protein &  0.004-0.04 &   \cite{Sen03JPCB} \\
\hline
\end{tabular}
\end{center}
\label{tab:delta}
\end{table}

Understanding, the dynamics of a quantum system which is strongly coupled to its environment is a challenging theoretical problem that has attracted considerable attention over the past few decades \cite{Weiss99}.
Many consider that decoherence is the key to resolving
the quantum measurement problem.\cite{Leggett05S,Schlosshauer04RMP,Zurek}
These issues are receiving renewed interest
because decoherence is detrimental
to quantum information processing.\cite{Unruh95PRA,Chuang95S}
 Substantial progress has been made by considering the simplest 
possible
models such as the spin boson model \cite{Weiss99}
 which describes a two-level system (the ``spin'')which is coupled linearly to an infinite ``bath'' of harmonic oscillators.

 For biomolecular systems, the spin-boson model has previously been
 applied to electron transfer.\cite{Xu94CP,Warshel01QRB,Garg85}
We have recently shown
the  relevance of the spin-boson model
to understanding the effect of the environment on
 F\"orster resonant energy transfer between two chromophores\cite{Gilmore06CPL}.
Of particular interest
 is the case where two molecules are coupled by
 Resonance Energy Transfer (RET), such as rings of chlorophyll molecules in photosynthesis and in Fluorescent Resonance Energy Transfer spectroscopy (FRET).  Here, an excitation in one chromophore may be transferred to a nearby chromophore by the Coulomb interaction, typically dipole-dipole interactions.  A coupled system of molecules such as this may be mapped to the spin-boson model \cite{Gilmore05JPCM},
 where the two quantum states
refer to the location of the excitation, $\epsilon$ is the difference in the two chromophore's excited energy levels, and $J(\omega)$ describes the coupling of the excitation to the environments surrounding each molecule. 
 We have previously shown 
\cite{Gilmore05JPCM} that the appropriate spectral density is simply the sum of the spectral density of each individual chromophore-protein complex.
  The magnitude of the spectral density
 then determines whether the transfer is coherent (oscillatory) or incoherent (one-way).  There are several definite experimental signatures of the coherent interaction of a pair of chromophores.  These include (Davydov) splitting of energy levels \cite{vanHolde98}, super- and sub-radiance (i.e., increase and reduction of the
 radiative lifetime\cite{Hettich02,Monshouwer97JPCB}) and changes in fluorescence anisotropy\cite{Kuhn02CP}. 
 Both coherence (within a ring) and incoherence (between rings)
 may play potentially
important functional roles in light harvesting complexes.

{\it The Hamiltonian.} For the oscillators this can be written as
\begin{equation}
H_0  = \sum_\beta \left( \frac{p_\beta^2}{2m_\beta} + \frac{1}{2} m_\beta \omega_\beta^2 q_\beta^2 \right)
\end{equation}
where $\beta$ is an index denoting a particular oscillator with
mass $m_\beta$, frequency $\omega_\beta$, momentum $p_\beta$, and position
$q_\beta$.
We introduce 
second quantised operators 
 $a_\beta$ and $a_\beta^\dagger$ 
where $ a_\beta  
\equiv 
 \sqrt{\frac{m_\beta \omega_\beta}{2 \hbar}}
(q_\beta + i p_\beta/(m_\beta \omega_\beta))$, and satisfy the boson
commutation relations 
 $[a_\beta,a_\gamma^\dagger] = \delta_{\beta,\gamma}$ 

The Hamiltonian of the whole system is
 \begin{equation}
H= H_0 + H_2
\end{equation}
where
\begin{equation}
H_2 = \left(\begin{array}{cc}\frac{\epsilon}{2}  +
 \sum_\beta C_\beta \sqrt{\frac{2m_\beta \omega_\beta}{\hbar}} q_\beta & \Delta \\\Delta & -\frac{\epsilon}{2} -
 \sum_\beta C_\beta
 \sqrt{\frac{2m_\beta \omega_\beta}{\hbar}}
 q_\beta\end{array}\right)
\end{equation}
In terms of the second quantised operators
this can be written
\begin{equation}
H = \frac{1}{2} \epsilon \sigma_z + \Delta \sigma_x + \sum_\beta \hbar\omega_\beta a_\beta^\dagger a_\beta + \sigma_z \sum_\beta C_\beta (a_\beta^\dagger + a_\beta),
\label{spin-boson}
\end{equation}
where 
$\sigma_x$
 and 
$\sigma_z$
 are Pauli spin matrices, the $C_\beta$ describes the coupling of the system to each bath mode $\beta$, $\epsilon$ is the separation of system energy levels and $\Delta$ is the tunneling matrix element coupling the two states. Another model is the spin-bath model \cite{Prokofev00RPP}, where the system of interest is coupled to specific localised states of the environment, themselves treated as two level systems.

{\it Spectral density.}
For the spin-boson model, the quantum dynamics
of the two-level system (TLS)
 is completely determined by a single function,
 the spectral density,\cite{Weiss99}
 which is defined by
\begin{equation}\label{eq:spectral}
J(\omega) = \frac{4\pi}{\hbar}\sum_\beta C_\beta^2 
\delta(\omega-\omega_\beta).
\end{equation}
It 
describes how strongly the oscillators with frequency near $\omega$ are coupled to the two-level system.
An important quantity is the {\it reorganisation energy}
defined by,
\begin{equation}\label{eq:reorg}
E_R = \int_0^\infty d\omega \frac{J(\omega)}{\omega}.
\end{equation}
 Many systems are described by {\it ohmic dissipation},
 for which $J(\omega) =\hbar \alpha \omega$, below some cutoff frequency,
 $\omega_c$, 
related to the relaxation rate of
the environment, and if $\Delta \ll \hbar \omega_c$ then
at frequencies higher than this cutoff
the coupling to the bath of oscillators can be neglected.
The main purpose of this paper is to derive physically realistic
expressions for this spectral density that are relevant to
biological chromophores interacting with their environment.

{\it Known results.}
 If $\epsilon,\Delta \ll \hbar\omega_c$, 
for ohmic dissipation $\alpha$ is a critical parameter for determining the qualitative properties
of the  quantum
 dynamics \cite{Weiss99,Lesage98PRL}. 
 At zero temperature, for $\alpha<\frac{1}{2}$ the state of the TLS,
 exhibits damped Rabi oscillations, a signature of quantum coherence
and interference.
This can be described by considering the time dependence of the
probability that the system is in one of the two levels,
which can be related to
the expectation value $\expec{\sigma_z(t)}$.
  For $\frac{1}{2} < \alpha <1$,
 the system exhibits incoherent relaxation (exponential decay of $\expec{\sigma_z(t)}$, and for $\alpha>1$ the system is localised in its initial state - an example of the quantum Zeno effect \cite{Joos84PRD}.
  A non-zero temperature reduces the range of $\alpha$ over which
 coherent oscillations can occur (see Fig. 21.2 in \cite{Weiss99}).

If $\Delta > \hbar\omega_c$ then the results of Refs. \cite{Weiss99} and \cite{Lesage98PRL} do not apply \cite{Carmeli88JCP}.  The bath responds slower than the relevant timescale for the 
dynamics of the TLS.
  Consequently, in order to destroy coherent oscillations, the bath must
 couple more strongly to the two-level system than for the case
 $\Delta \ll \hbar\omega_c$.
  System dynamics has been studied with quantum  
 Monte Carlo simulations\cite{Muhlbacher03JCP} 
 when $\epsilon = 0$.
Coherent oscillations in $\expec{\sigma_z(t)}$ may be present
 for $\alpha>1$.   Fig. 13 in
 Ref. \cite{Muhlbacher03JCP} shows that when $\Delta=6\hbar\omega_c$
 then coherent oscillations can exist even for $\alpha=30$. 
 Fig. 8 of 
Ref. \cite{Bulla05PRB} shows that for $\Delta=\hbar\omega_c$,
numerical renormalisation group calculations predict that
 coherent oscillations exist for $\alpha<1.5$.  
Fig. 3 of Ref. \cite{Kehrein96ZPB}
shows that a renormalisation flow equation
approach predicts that
when $\Delta/\hbar\omega_c$ increases from very small
values to 0.3 that the allowed parameter range 
for coherent oscillations increases to $\alpha < 1.5$.
The coherent-incoherent transition is associated with the
delocalisation-localisation transition that has been studied
widely in chemistry in systems 
such as the Creutz-Taube ion and the special pair in photosynthetic
systems.\cite{Reimers96CP}
 
The quantum dynamics of the spin boson 
model \eqref{spin-boson} for a general spectral density $J(\omega)$ will be largely determined by the magnitude and frequency dependence of $J(\omega)$ for $\omega\sim \Delta$.
 For example, when the bath is weakly coupled 
(i.e., $J(\Delta) \ll \Delta$)
to an unbiased ($\epsilon=0$) two level system 
 coherent oscillations exist, and the relevant decoherence rate given
by Fermi's Golden Rule is\cite{Weiss99} 
 \begin{equation}
\frac{1}{T_2} = J(\Delta) \coth \left(\frac{\Delta}{2k_B T}\right).
\end{equation}

\subsection{The chromophore environment: protein, bound water,
 and bulk water}

The structures and dynamics associated with the interaction of proteins with water is extremely rich and a challenge to model and to
 understand.\cite{Edsall83AP,Wand01NSB,Pal04CR,Niimura05}.
  One can classify the water molecules associated
with proteins into several categories.
(i) Water which is distant from the protein and has the
same properties as bulk water.
(ii) Water at the surface of the protein molecule.
The first layer of molecules is referred to as the first hydration
or solvation layer. These molecules are weakly bound to the
charged residues found at the protein surface.
(iii) Water buried inside the protein and which often
binds to specific sites in the protein via
multiple hydrogen bonds.
The water inside and at the surface of the protein can exchange with the bulk water.

 Advances in experimental probes such as neutron
 scattering\cite{Niimura05}, nuclear magnetic resonance
 (NMR)\cite{Wand01NSB},
 femtosecond laser spectroscopy,\cite{Pal04CR}
 and dielectric dispersion \cite{Grant78} has allowed a quantitative description of the properties
 of the water molecules associated with specific parts
of the solvated protein.
 Key quantities that can be determined include,
(a) the occupancy (i.e., the probability that a water
molecule will be found at the site),
(b) the residence time (the timescale for exchange of
the water molecule with the surrounding bulk water), and (c)
the ``order parameter'' which is a measure
of the rotational freedom of the water molecule
at the site.
NMR measurements suggest that the molecules at the
surface exchange with the bulk water on timescales
ranging from 10 psec to 1 nsec\cite{Wand01NSB}.
In contrast, buried molecules exchange with the solvent
on timescales of the order of 1 ns to 1 $\mu$s \cite{Wand01NSB}.

The term ``biological water'' has been used to describe water in proximity to a biological macromolecule \cite{Nandi97JPCB}. Dielectric relaxation is significantly different in biological water \cite{Bhattacharyya03ACR}.  Whereas in bulk water the dominant dielectric relaxation time is
8.3 ps, for bound water this can be 2 to 4 orders of magnitude larger. Dielectric spectroscopy measurements of proteins in aqueous solutions found four dielectric relaxation times.\cite{Grant78}
For myoglobin these times were attributed to
(1) reorientation of bulk water (8 ps),
(2) relaxation of water associated with the protein
(10 ps and 150 ps), and (3) reorientation of the whole
protein molecule (15 nsec).

Given the structural, chemical, and dynamical complexity of these
environments we briefly discuss the limitations of some of
the underlying simplications and approximations we assume in our models.
Although, these simplifications may lead to quantitative differences
between the predictions of our models and real systems we do
not anticipate qualitative differences.

{\it Spherical symmetry.}
We assume that the chromophore is located at the centre of a spherical
cavity inside a spherical protein. 
Similar assumptions have been made in some other
studies of dielectric relaxation in proteins.\cite{Hofinger01,Simonson01}
It has been found\cite{Hsu97JPCB} that
there are only small quantitative differences between the dielectric
relaxation associated with elliptical cavities compared to spherical ones.
Clearly our results will be most relevant to globular proteins
with a chromophore towards the centre.
A more serious concern is that some chromophores can be located near the
protein surface and so more exposed to the solvent.
They may be better modelled by a spherical vacuum cavity at the
planar interface between two different dielectric medium.
Similar geometric considerations apply to 
transmembrane photosynthetic proteins and systems
containing lipid membranes.
The approach used here could be extended to such cases
by considering continuum dielectric models with
different geometries.\cite{Cramer99CR,Song98JCP}

{\it Point dipole approximation.}
The chromophore is treated as a point dipole and its spatial extent
is neglected. More realistic treatments of spatially extended
distributions inside solvent cavities,\cite{Song98JCP}
do not lead to qualitatively different behaviour.

{\it Neglect of the nuclear dynamics of chromophore.}
Most electronic transitions will be associated with some
structural change of the chromophore. Except, for the case of
chromophores (such as retinal and PYP) which undergo
conformational change upon photo-excitation, generally the
reorganisation energy (and associated Stokes shifts) 
for modes with frequencies less than 1000 cm$^{-1}$ are
typically of order of 10's cm$^{-1}$ and so
much smaller than those associated with the solvent
and protein.\cite{Jimenez02JPCB}.
Furthermore, intra-molecular vibrations with substantial
reorganisation energies have  sufficiently
high frequencies that they occur on timescales
much faster than most of the experiments we consider and
are not thermally excited at room temperature.

\subsection{Overview of the paper}
\label{overview}

In this paper we consider five distinct dielectric
continuum models of the environment of a biological
chromophore. For each model
 we derive an expression for the spectral density \eqref{eq:spectral}.
 This allows us to explore how the relative importance
 of the dielectric relaxation of
the solvent, bound water, and protein depends
 on the relevant length scales (the relative size of the chromophore, the protein and the thickness of the layer of bound water) and time scales (the dielectric relaxation times of the protein, bound water and the solvent) as is discussed in Section \ref{sec:optical spectroscopy}. Many experimentally obtained spectral densities
 can be fitted to a sum of Lorentzians 
(see Table \ref{tab:exp data}).
For a protein that is large compared to the size of the binding pocket of the chromophore and the width of the bound water layer, our models predict a spectral density given by the sum of three Lorentzians which correspond to the dynamics of the protein, bound water and bulk water dynamics respectively.  
An essential feature is the separation of time scales,
associated with the solvation coming from each of the three
components of the environment.

Specifically, to a good approximation, the spectral density is given by,
\begin{equation}
J(\omega) = \frac{\alpha_p \omega}{1+(\omega\tau_p)^2} + \frac{\alpha_b \omega}{1+(\omega\tau_b)^2} + \frac{\alpha_s\omega}{1+(\omega\tau_s)^2}.
\label{eq:3J}
\end{equation}
The subscripts $x=p,s,b$ refer to the protein, solvent
 and bound water, respectively.
We show that when the dielectric relaxation of
the different components of the environment are treated
in a Debye approximation, that
the  relaxation times can be expressed as:
\begin{equation}
\label{taup}
\frac{\tau_p}{\tau_{D,p}} = \frac{2\epsilon_{p,i}+1}{2\epsilon_{p,s}+1}
\end{equation}
\begin{equation}
\label{taus}
\frac{\tau_s}{\tau_{D,s}} = \frac{2\epsilon_{s,i}+\epsilon_{p,i}}
                {2\epsilon_{s,s}+\epsilon_{p,i}}
\end{equation}
\begin{equation}
\label{taub}
\tau_b = \tau_{D,b}
\end{equation}
where $\epsilon_{x,s},\epsilon_{x,i},\tau_{D,x}$ are, respectively,
 the static dielectric constant, high frequency
dielectric constant, and relaxation times of a
 Debye model for each medium, $x=p,s,b$. 
The high frequency dielectric constant is related to the refractive index $n_x$, by
$\epsilon_{x,i}=n_x^2$.

We show that the
 reorganisation energies associated with each part of the 
environment are
 given by $\alpha_x/\tau_x$, where
\begin{equation}
\frac{\alpha_p}{\tau_p} = \frac{3 (\Delta\mu)^2}{\pi\epsilon_0 a^3} \frac{  (\epsilon_{p,s}-\epsilon_{p,i})}{(2\epsilon_{p,s}+1)(2\epsilon_{p,i}+1)}
\end{equation}
\begin{equation}
\label{alphas}
\frac{\alpha_s}{\tau_s} = \frac{3(\Delta\mu)^2}{\pi\epsilon_0 b^3} \frac{ (\epsilon_{s,s}-\epsilon_{s,i})}{(2\epsilon_{s,s}+
\epsilon_{p,i})(2\epsilon_{s,i}+
\epsilon_{p,i})} \left(\frac{9\epsilon_{p,i}}{(2\epsilon_{p,i}+1)^2}\right)
\end{equation}
\begin{equation}
\label{alphab}
\frac{\alpha_b}{\tau_b} = \frac{3 (\Delta\mu)^2}{2\pi\epsilon_0 b^3} \left( \frac{c-b}{b} \right) \frac{(\epsilon_{b,s}^2+ 2\epsilon_{s,s}^2 )
 (\epsilon_{b,s}-\epsilon_{b,i})}
{\epsilon_{b,s}^2 ( 2\epsilon_{s,s}+\epsilon_{p,i})^2}
\end{equation}
In particular, for typical systems the above three quantities can be of the same order of magnitude, i.e.,
\begin{equation}
\label{eq:compare}
\frac{\alpha_s}{\tau_s}\sim \frac{\alpha_b}{\tau_b}
 \sim \frac{\alpha_p}{\tau_p}.
\end{equation}
Due to the large separation of time scales, the spectral
density (\ref{eq:3J}) will have peaks at approximately,
$\omega = 1/\tau_x$. 
Hence, the peaks  are
of approximate magnitude, $J(\omega=1/\tau_x) \sim \alpha_x/\tau_x$,
and can be of the same order of magnitude.
 This is because although
each contribution
 involves  different dielectric constants,
they only have a limited range of values, and the ratios of
the different dielectric constants that appear
on the right hand side of the above expressions 
are all of order one.
 This is supported by experimental data (see Table \ref{tab:exp data})
  where several relaxation times are observed which vary by several orders of magnitude, but whose relative contributions are
of comparable magnitude.
 Hence, in many cases only a single component of
 the environment (protein, bound water, or bulk solvent)
 will be relevant to a given process.

{\it The effect of the protein.}
 The expression (\ref{alphas})
 allows us how the ultra-fast solvation associated
with the solvent is modified
in the presence of a protein. Water is a highly polar medium with
$\epsilon_{s,s} \simeq 80$,
and $\epsilon_{s,i} \simeq 4$,
and fast dielectric relaxation,
$\tau_{D,s} \simeq 8$ psec.
Hence,
we see that even if 
$\epsilon_{p,i}$ is as large as
5 that the solvation time associated with the solvent
is only  be 50 per cent larger compared
to the time of 0.3 psec which occurs in the absence of the protein.
The effect of the protein on the strength of the
coupling of the chromophore to the solvent is more substantial.
In (\ref{alphas}), the coupling scales with 
the inverse cube of the power of the radius of the protein.
Hence, if the diameter of the protein is four times the size
of the chromophore, the coupling $\alpha_s$ will be
reduced by two orders of magnitude.
The above results show that even a distant
solvent can lead to ultra-fast solvation comparable to
that found in the absence of the protein.
This leads us to suggest that some studies which claim
to have identified
 ultra-fast dielectric relaxation of proteins \cite{Homoelle98JPCB,Riter96JPCB} may in fact be detecting the fast response of the distant solvent.
 

{\it Order of magnitude estimates of $\alpha_x$.}
    Typical values of parameters  are
 $a \sim 3 \AA$,
 $b \sim 10 \AA$,
 $\Delta \mu \sim 1 D$, and so the reorganisation
energy $\alpha_x \hbar / \tau_x \sim 10$ meV $\sim 100 $ cm$^{-1}$.
From Figure 2 and the Table in the Appendix
we see that $\hbar / \tau_x $ is of 
the order of 10 meV, 1 meV, and 0.01 meV, for the solvent,
bound water, and protein, respectively.
Hence, the dimensionless couplings 
 $\alpha_s \sim 1$,
 $\alpha_b \sim 10$,
 and $\alpha_p \sim 100$.
Hence, the only quantum dynamics that is likely to
be coherent is that which occurs on timescales comparable to
or faster than the relaxation of the bulk water,
i.e., less than a picosecond.

The outline of the paper is as follows. 
 In Section \ref{sec:sbm} we describe how the interaction between a chromophore and its protein and solvent environment may be modelled by an
independent boson model.  We show how the interaction
with the environment leads to decoherence of
the electronic states of the chromophore.
In Section \ref{sec:continuum} we 
propose a set of continuum dielectric models suitable for describing the environment around a chromophore, and use them to obtain an expression for the spectral density in each case.  In Section \ref{sec:evaluating} we consider particular limits of these spectral densities and obtain simple expressions for the contribution of each component
of the environment (protein, bound water, and bulk water)
 to the total spectral density.
In particular, we are able to obtain expressions that can
be used to evaluate the relative importance of
each component of the environment.
We find that even when the chromophore is 
completely surrounded by a protein it is possible
that the ultra-fast solvation (on the psec timescale)
is dominated by the bulk solvent surrounding
the protein.
In Section \ref{sec:optical spectroscopy} we discuss methods for obtaining spectral densities from optical spectroscopy, and compare the predictions of our models to experimental data.
In Section \ref{sec:simulations}
we relate our results for the spectral density due to dielectric
relaxation to what has been learnt from molecular dynamics simulations
on specific protein systems.

\section{Quantum dynamics of the Independent
boson model 
}\label{sec:sbm}

It can be shown \cite{Gilmore05JPCM} that the coupling of the electronic excitations in a chromophore to its environment may be modelled by an independent boson model \cite{Mahan90}, which has the Hamiltonian,
\begin{equation}
\label{eq:indep boson}
H = \frac{1}{2} \epsilon \sigma_z + \sum_\beta \omega_\beta 
 a_\beta^\dagger a_\beta  +  \sigma_z \; \sum_\beta  C_\beta
  (a_\beta +a_\beta^\dagger).
\end{equation}
We note that this is just the spin boson model with $\Delta=0$.
Here the chromophore is treated as a two level system with energy gap $\epsilon$ between the ground and excited state.  The first term describes the energy of the isolated chromophore, described by Pauli sigma matrix $\sigma_z$.  The second term is the energy of the surrounding environment (protein and solvent), where the environment is modelled as a bath of harmonic oscillators \cite{Gilmore05JPCM}. 
 The final term describes the coupling of the state of the chromophore $\sigma_z$ to the environment.
  In this case, the coupling \cite{Onsager36JACS,Bottcher73} is due to electrostatic interactions between the chromophore dipole and the ``cage'' of polarised solvent and protein molecules around it, as will be described in more detail below. 
The effect of this coupling on the quantum dynamics of the chromophore is completely specified by the spectral density,
 defined by eq. (\ref{eq:spectral}) 
\cite{Mahan90}. 

If the two-level system is initially ($t=0)$
in a coherent superposition state $|\Psi> = a|1> + b|2>$,
which is not coupled to the environment.
An ultra-fast laser pulse can create such a state.
Then at time $t$, the TLS is described 
by a $2\times 2$ reduced density matrix, $\rho(t)$.
It has matrix elements \cite{Reina02} 
\begin{align}\label{density}
\rho_{11}(t) &= \rho_{11}(0)= |a|^2 \-
\rho_{22}(t) &= \rho_{22}(0) = |b|^2= 1-\rho_{11}(0) \-
\rho_{12}(t) = \rho_{21}^\ast(t) &= a^*b 
\exp(-i\epsilon t + i\theta(t)-\Gamma(t,T))
\end{align}
where $\theta(t)$ is a phase shift given by
\begin{equation}\label{eq:theta}
\theta(t) = \int_0^\infty 
 d\omega
J(\omega) \frac{\left[\omega t-\sin(\omega t)\right]}{\omega^2}
\end{equation}
and
\begin{equation}
\Gamma(t,T) = \int_0^\infty d\omega J(\omega) \coth\left(\frac{\omega}{2k_B T}\right) \frac{(1-\cos\omega t)}{\omega^2}
\end{equation}
describes the decoherence due to interaction with the environment.

The phase shift can be used to define a time dependent 
Stokes shift of the energy separation of the two
levels. The instantaneous energy is
found by taking the derivative of
\eqref{eq:theta} with respect to time,\cite{Fleming96,Hsu97JPCB}  
\begin{equation}
\label{eq:nut}
\nu(t)=\epsilon-\frac{d\theta(t)}{dt}=
\epsilon-E_R -
  \int_0^\infty d\omega
\frac{J(\omega)}{\omega} \cos(\omega t)
\end{equation}
Hence, the 
spectral density can be determined by taking the
Fourier transform of measurements
of $\nu(t)$, as  discussed in Section \ref{sec:optical spectroscopy}.

Depending on the relative size of the time $t$ to the time scales defined by $1/\omega_c$ (the relaxation time of the bath)
and $\hbar/k_B T$, there are three different regimes of time dependence.  For short times $\omega_c t < 1$,
\begin{equation}
\Gamma(t, T) = \frac{t^2}{2\tau_g^2}
\end{equation}
and so there is a Gaussian decay of coherence,
on a time scale $\tau_g$, given by
\begin{equation}
\label{eq:gaussian}
\frac{1}{\tau_g^2} = \int_0^\infty d\omega J(\omega) \coth\left( \frac{\omega}{2k_B T}\right).
\end{equation}
If in addition,  $k_B T \gg \hbar\omega_c$, this reduces to
\begin{equation}
\frac{\hbar}{\tau_g} = \sqrt{2 E_R k_B T}
\end{equation}
where $E_R$ is the reorganisation energy given by
Eqn. (\ref{eq:reorg}).

For intermediate times,
$1/\omega_c < t < \hbar/k_B T$ (the quantum regime \cite{Unruh95PRA};
it only exists if
$k_B T < \hbar\omega_c$), 
then
\begin{equation}
\Gamma(t,T) \approx \alpha \ln (\omega_c t),
\end{equation}
where $\alpha \equiv J'(\omega=0)$,
leading to a power law decay of coherence.

For long times ($t \gg \hbar/k_B T$, the thermal regime)
\begin{equation}
\Gamma(t,T) \approx 2 \alpha  k_B T t/\hbar \equiv t / \tau_d,
\end{equation}
giving exponential decay of coherence.

The crossover from Gaussian to exponential decay
has been explored in quantum measurement theory in the context of continuous measurement and the quantum Zeno effect \cite{Joos84PRD}.
The decay of the off-diagonal part of the density matrix
results from decoherence from the interaction of the
TLS with the environment. 
Thus, we see the effect of the environment on the TLS is
\begin{equation}
\rho(t=0) = 
\left(\begin{array}{cc}
|a|^2 & a^* b \\
a b^* &  |b|^2
\end{array}\right)
 \ \ \ \rightarrow \rho(t \to \infty)
=\left(\begin{array}{cc}
|a|^2 & 0 \\
0 &  |b|^2
\end{array}\right)
\end{equation}
Hence, the timescale $\tau_d$
 can be interpreted as the ``collapse'' of the wavefunction of
the TLS due continuous measurement of the state of the
TLS by the environment.\cite{Joos84PRD,Schlosshauer04RMP,Zurek}
We will see that using the spectral densities that we extract from
experiment and from our continuum dielectric models that typically
$\alpha > 1$, and so at room temperature, $\tau_d < 10$ fsec.

\section{The spectral density for the different
 continuum models of the environment}\label{sec:continuum}

In the simplest continuum model \cite{Hofinger01} picture of protein-pigment complexes, 
 the chromophore can be treated as a point dipole inside a spherical dielectric \cite{Hofinger01,Bottcher73}, representing a
globular protein, surrounded by a uniform polar solvent with complex dielectric constant $\epsilon_s(\omega)$ \cite{Onsager36JACS}.  This can also apply to a chromophore-protein complex embedded in a solid dielectric medium.  In a previous work \cite{Gilmore05JPCM}, the spectral density was determined for a free chromophore in a solvent. 
 However, many
chromophores are inside proteins, which may have a significant effect
on the coupling to the environment,
 if only in pushing back the solvent.  

\begin{figure}[htbp]
\begin{center}
\includegraphics[width=3.5in]{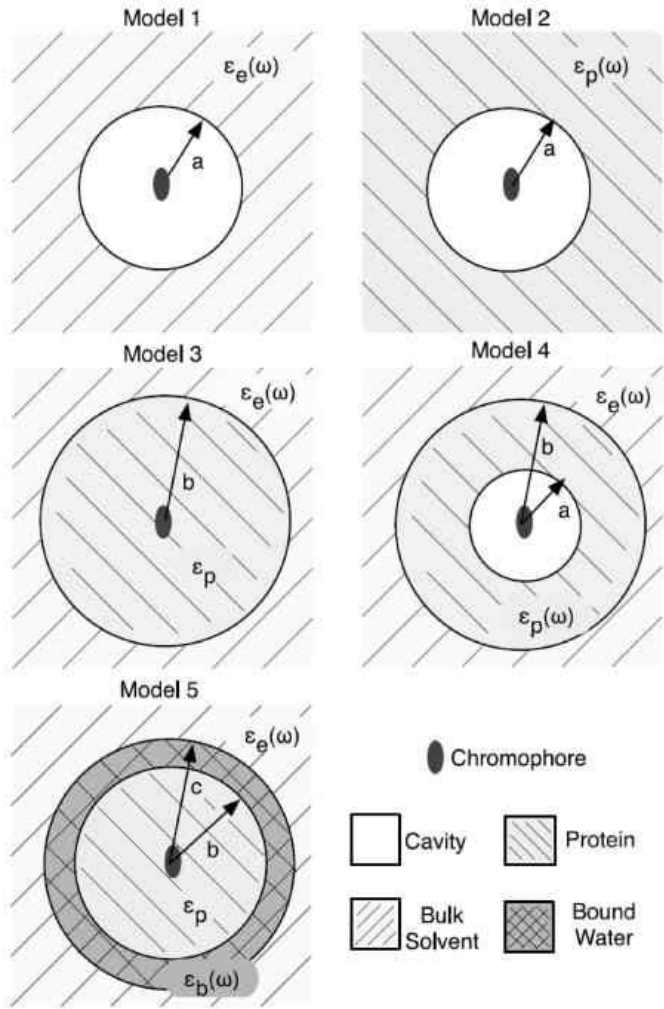}
\caption{The five continuum
dielectric 
models considered for a chromophore-protein-solvent system.  The chromophore is modelled as a point dipole.
  In Model 1, the chromophore is modelled  sits at the centre of a cavity of radius $a$ roughly of the van de Waals size of the chromophore, surrounded by a uniform polar solvent with complex dielectric constant $\epsilon_s(\omega)$.  In Model 2, the chromophore is surrounded by an
 infinite protein, modelled as a uniform, continuous dielectric
 medium, with complex dielectric constant $\epsilon_p(\omega)$.  In Model 3, the chromophore sits in a protein of radius $b$ surrounded by the solvent.  In Model 4, the chromophore sits in a cavity inside the dynamic protein, surrounded by solvent.  In Model 5, the static protein is surrounded by a thin shell of bound water of radius $c$, surrounded by the bulk solvent.}
\label{fig:5models}
\end{center}
\end{figure}

Several continuum dielectric
models for the protein environment have
previously been proposed (see for example Figure 2 in ref. \cite{Hofinger01}  and Figure 2 in ref. \cite{Voges98}.)  These models often provide good qualitative and rough quantitative agreement \cite{Simonson01}.   We consider five distinct models, illustrated in figure \ref{fig:5models}. In every case, the central chromophore dipole polarises the surrounding cage of protein and solvent, which in turn produces an electric field
 inside the cavity, called the \emph{reaction field} \cite{Onsager36JACS, Bottcher73}.  The interaction
 of this field with the central dipole is responsible for the interaction between the chromophore and its environment. The independent
boson model can be obtained by writing the reaction field in
 terms of its normal modes, and quantising the coefficients in the standard second quantisation method \cite{Gilmore05JPCM}, and the fluctuation dissipation theorem used to relate fluctuations in the reaction field to the appropriate spectral density \cite{Gilmore05JPCM}. A detailed derivation applicable to all models considered below, is
given in the Appendix.
  
  {\bf Model 1.}
This describes a free chromophore with no surrounding protein.
  The molecule sits inside
 a spherical cavity of radius $a$, approximately the van de Waals
 radius of the molecule, inside a solvent with dielectric constant
 $\epsilon_s(\omega)$.  The spectral density is\cite{Hsu97JPCB}
\begin{equation}\label{eq:J(omega) solvent only}
J_1(\omega) =  \frac{(\Delta\mu)^2}{2\pi\epsilon_0 a^3} \Im \frac{\left(\epsilon_s(\omega)-\epsilon_c \right)}{2\epsilon_s(\omega)+\epsilon_c}.
\end{equation}
where $a$ is the radius of the cavity containing the chromophore
 and $\epsilon_s(\omega)$ is the complex dielectric function of the solvent and $\epsilon_c$ the (static) dielectric constant of the cavity. $\Delta\mu$ is the difference
between the dipole moment of the chromophore 
in the ground and excited states.

  {\bf Model 2.}
This is  somewhat
 analogous situation to Model 1, but in this 
case the chromophore is surrounded by an infinite,
 uniform protein with complex dielectric constant $\epsilon_p(\omega)$.
This chromophore
is  again inside a cavity of radius $a$, which would be approximately
 the same radius as for Model 1. Such a model may be appropriate when the
 protein is very large.  The spectral density has a similar
form to Model 1, except involves
 the dielectric constant of the protein $\epsilon_p(\omega)$:
\begin{equation}\label{eq:J(omega) protein only}
J_2(\omega) =  \frac{(\Delta\mu)^2}{2\pi\epsilon_0 a^3} \Im \frac{\left(\epsilon_p(\omega)-\epsilon_c \right)}{2\epsilon_p(\omega)+\epsilon_c}.
\end{equation}
 
{\bf Model 3.}
The chromophore is surrounded by a uniform dielectric sphere representing the protein. This protein sphere has radius $b$ and (constant) dielectric constant $\epsilon_p$.   The spectral density is
\begin{equation}\label{eq:J(omega) uniform protein}
J_3(\omega) =  \frac{(\Delta\mu)^2}{2\pi\epsilon_0 b^3}
 \Im \frac{\left(\epsilon_s(\omega)-\epsilon_p \right)}{2\epsilon_s(\omega)+\epsilon_p}.
\end{equation}
where $\epsilon_s(\omega)$
 is the complex dielectric function of the solvent.

 {\bf Model 4.}
In a more detailed picture,
 (see the Figure  the Appendix), Models 2 and 3 are combined so that the chromophore sits inside a hollow cavity within the protein.  The cavity has radius $a$ (typically the size of the chromophore) and vacuum dielectric constant $\epsilon_0$, while the protein again has radius $b$ and is now described by complex dielectric constant $\epsilon_p(\omega)$.  Further detail may be added by treating the outer layer of the protein sphere as a separate, higher dielectric medium \cite{Hofinger01} representing the charged surface groups \cite{Schutz01}.  In all cases, the chromophore is treated as a point dipole. 
 The spectral density 
 is given by
\begin{equation}\label{eq:J(omega) model 4}
J_4(\omega) = \frac{(\Delta\mu)^2}{2\pi\epsilon_0 a^3} \Im
\frac{ (\epsilon_p+2)(\epsilon_s-\epsilon_p) a^3 + (\epsilon_p-1)(2\epsilon_s + \epsilon_p) b^3}{ 2(\epsilon_p-1)
(\epsilon_s-\epsilon_p) a^3 + (2\epsilon_p+1)(2\epsilon_s+\epsilon_p) b^3}
\end{equation}
where $b$ is the radius of the protein containing the chromophore, $a$ is the radius of the cavity containing the chromophore (usually the size of the chromophore), $\epsilon_s$ is the complex dielectric function of the solvent and $\epsilon_p$ is now the complex dielectric function of the protein. Note that the frequency dependence of the dielectric constants has been omitted for clarity.  We see that for appropriate limits ($(a/b) \rightarrow 0$, $\epsilon_p=\epsilon_s$, $\epsilon_p=1$, etc) Models 1 and 2 can be recovered, as expected.  (To obtain Model 3, one would have to allow the centre cavity in Model 4 to have an arbitrary dielectric constant.)

{\bf Model 5.}
The chromophore sits in a static protein (no cavity)
 and is surrounded by a thin shell of bound water with a different dielectric constant to the bulk solvent. To obtain the spectral density,
 we can use the results for Model 4, with $\epsilon_p \rightarrow \epsilon_b, \epsilon_c \rightarrow \epsilon_p, a\rightarrow b$,
and $b\rightarrow c$:
\begin{equation}\label{eq:J(omega) model 5}
J_5(\omega) = \frac{(\Delta\mu)^2}{2\pi\epsilon_0 b^3} \Im
\frac{ (\epsilon_b+2\epsilon_p)(\epsilon_s-\epsilon_b) b^3 + (\epsilon_b-\epsilon_p)(2\epsilon_s + \epsilon_b) c^3}{ 2(\epsilon_b-\epsilon_p)(\epsilon_s-\epsilon_b) b^3 + (2\epsilon_b+\epsilon_p)(2\epsilon_s+\epsilon_b) c^3}
\end{equation}
Note that $\epsilon_p$ refers to a constant (typically high frequency) protein dielectric viz. Model 2.  $\epsilon_b$ is the complex dielectric of the bound water.

\subsection{Debye model for the frequency dependence of 
the dielectric constants}

To specify the dielectric constant of each component 
of the environment $\epsilon_x(\omega)$ ($x=s,p,b$)
we consider the Debye form of the dielectric constant\cite{Hsu97JPCB},
\begin{equation}\label{eq:Debye dielectric}
\epsilon_x(\omega) = \epsilon_{x,i} + \frac{\epsilon_{x.s}-\epsilon_{x,i}}{1-i\omega\tau_{D,x}}
\end{equation}
where $\tau_{D,x}$ is the Debye relaxation time of the 
relevant component,
$\epsilon_{x,s}=\epsilon(\omega=0)$ is the static dielectric constant, and $\epsilon_{,i}=\epsilon(\omega\rightarrow \infty)$ is the high frequency dielectric constant, within the range of physically relevant frequencies --  note that $\epsilon_{x,i}$ would always be $1$ for sufficiently high frequencies \cite{Song93JCP}. 
For water at room temperature,  $\epsilon_{s,s}=78.3$, $\epsilon_{s,i}
=4.21$ and $\tau_{D,s} = 8.2$ ps \cite{Afsar78}.
For comparison, for
 THF (tetrahydrofuran) the
values  are $\epsilon_{s,s}=8.08$, $\epsilon_{s,i}=2.18$ and
 $\tau_{D,s}=3$ ps \cite{Horng95JPC}.

In Model 4, the protein is treated as a complex frequency dependent dielectric with dielectric constant $\epsilon_p(\omega)$, and as such is allowed to relax and respond to the chromophore.  We will consider the case of a Debye dielectric \cite{Song93JCP}, but a more complicated model including multiple relaxation times is also possible \cite{Kindt96JPC}.  Typical values for $\epsilon_p$ are between 4--40 depending on which part of the protein is of
 interest \cite{Pitera01,Hofinger01,Loffler97}.
  Studies also suggest that charged groups on the surface of the protein can skew the average value of the protein dielectric, and it may be more appropriate to model such proteins as having an inner and an outer shell with two different dielectric constants.  The high frequency constant, which is the only value used by many studies, is more difficult to determine but is generally assumed to be between 1.5 and 2.5 \cite{Moog84B,Monshouwer97JPCB}.
Section \ref{sec:simulations}
 discusses determinations of $\epsilon_p(\omega)$
from molecular dynamics simulations.

The appropriate dielectric relaxation time of the protein may be different from the protein relaxation times, which can be of order milliseconds \cite{vanHolde98}, as there are processes (e.g., vibration of bonds) on the order of femtoseconds \cite[p132, Table 3.13]{vanHolde98} which may contribute to the dielectric function. These may be missed in studies of the dielectric constant on the nanosecond timescale \cite{Trissl01B} and is perhaps unobservable for aqueous solutions of proteins (e.g., ref. \cite{Pal04}).  Molecular dynamics simulations \cite{Loffler97} suggest a protein dielectric relaxation time of 10 ns for a peptide, while vibrations may be of the order of 100 fs which may apply in certain situations. Other studies have found no single
relaxation times, with relaxation processes occurring across the entire experimental range of 20 ps -- 20 ns\cite{Pierce92JPC,Pal01JPCB}.

For Model 3, an appropriate value for the constant dielectric of the protein must be chosen, which will depend on the frequency range of physical interest.  For example, for frequencies greater than $1/\tau_p$, where $\tau_p$ is approximately the protein dielectric relaxation time, the protein will be well approximated by its high frequency value.

The hydration shell of hydrogen bonded water molecules surrounding the protein 
will have a dielectric constant different to that of the bulk solvent,
and have a longer Debye relaxation time. 


\section{Analysis of Model 4 and Model 5}\label{sec:evaluating}

\subsection{Model 4 - Dynamic protein and dynamic solvent}
\label{sec:Model 4}

The full spectral density (Model 4) describes the total coupling to the protein and solvent, but is too complex to be written analytically and because of the multiple timescales involved may be non-Ohmic,
in certain frequency ranges.
However, there are many cases where the use of one of the simpler descriptions (Models 1-3) would be preferable. For example if the frequency
dependent dielectric constant 
of either the protein or solvent were not known, it 
would be useful to have a simple
criteria to
establish  whether obtaining these parameters is worthwhile.
If protein contributes negligible coupling beyond pushing the
 solvent back to a new distance, then specially obtaining its dielectric through experiment or simulation would be unnecessary.

 If either the solvent or the protein can be deemed unimportant, then simulations of the chromophores do not require their inclusion, saving valuable computational power. Conversely, if (for example) the solvent can be shown to have a significant effect on the chromophore, then treating the protein only will be insufficient and solvent effects must be included.
Hence, we explore under what conditions the protein dynamics may be neglected. The following discussion assumes that the dielectric relaxation time of the protein is longer than that of the solvent, which is expected to be true in the vast majority of cases.

\subsubsection{Limit of a small chromophore surrounded by a large protein}\label{sec:small cavity}

In the case of an ``infinite'' protein $a/b \rightarrow 0$, the full spectral
 density (Model 4) reduces to $J_2(\omega)$,
 the spectral density for Model 2.
In this limit, the solvent can be neglected when compared to the protein.  We might naively assume then that, since the ratio $a/b$ appears cubed,
 when the protein is several times larger than the chromophore the above
 spectral density can be used and the coupling 
to the protein is far stronger than to the solvent.
  However, a closer examination of \eqref{eq:J(omega) model 4} shows that if $\epsilon_p(\omega)\approx 1$, the solvent will be more significant even for large values of $b/a$, and a more appropriate expression for the spectral density describes the chromophore in a cavity of radius $b$ surrounded by solvent.
Hence, we seek a more precise statement for when the protein can be ignored.  The spectral density  can be rewritten:
\begin{equation}
J_4(\omega) = \frac{(\Delta\mu)^2}{2\pi\epsilon_0 a^3} \frac{ \left(\frac{\epsilon_p+2}{2\epsilon_p+1}\right) \left(\frac{\epsilon_s-\epsilon_p}{2\epsilon_s+\epsilon_p}\right) \frac{a^3}{b^3} + \left(\frac{\epsilon_p-1}{2\epsilon_p+1}\right)} {\left(\frac{2(\epsilon_p-1)}{2\epsilon_p+1}\right) \left(\frac{\epsilon_s-\epsilon_p}{2\epsilon_s+\epsilon_p}\right) \frac{a^3}{b^3} +1}
\end{equation}
We can  expand this expression in $\frac{\epsilon_s-\epsilon_p}{2\epsilon_s + \epsilon_p} \frac{a^3}{b^3}$.  Note that although this is a complex quantity, provided both the real and imaginary parts are small compared to unity we can use the Taylor series expression $\frac{ax+b}{cx+1} \approx b + (a-bc)x$.  Furthermore, both the real and imaginary parts of the prefactor of this expansion coefficient will  be less than unity, and a sufficient condition for this expansion to be valid is that $a/b$ be small. We find
\begin{align}\label{eq:taylor}
J_4(\omega) &\approx J_2(\omega) + \frac{(\Delta\mu)^2}{2\pi\epsilon_0 b^3} \Im \left[\frac{\epsilon_s(\omega)-\epsilon_p}{2\epsilon_s(\omega)+\epsilon_p}\right] \left(\frac{9\epsilon_p}{(2\epsilon_p+1)^2}\right) \-
&= J_2(\omega) + J_s(\omega)
\end{align}
where $J_s(\omega)$ represents the solvent contribution to the spectral density.  Note that the dynamics of the protein only contribute to the first term, which conversely contains no reference to the solvent. The second term includes the solvent dynamics plus the high frequency limit of the protein dielectric constant as the only relevant protein property.
 Thus, we identify the first term with the protein contribution and the second term with the solvent, modified by the presence of the protein's high frequency dielectric.

\subsubsection{Relevance of protein vs. solvent}

By comparing the magnitudes of these two terms in \eqref{eq:taylor}, we can establish the relative importance of the solvent and protein over different frequency ranges. We might expect that around $\omega = 1/\tau_p$ and $\omega=1/\tau_s$ that the protein and solvent
contributions should dominate respectively. Therefore, there should be a crossover point where the two contributions are roughly equal, somewhere in the range $1/\tau_p \ll \omega \ll 1/\tau_s$, which is what we look for now.

As defined above, the cut-off frequency of $\omega_p = 1/\tau_p$. Assuming a Debye dielectric for the protein, eq. 
\eqref{eq:J(omega) protein only}
 can be approximated to first order in $\omega_p/\omega$ as:
\begin{equation}
J_2(\omega) \approx \omega_p^2 \alpha_p / \omega, \qquad \omega \gg \omega_p
\end{equation}
Therefore, the tail end of the spectral densities given fall off as $1/\omega$ (as compared to their linear rise for $\omega\ll\omega_p$).
Noting that the reorganisation energy for this spectral density is $E_p = \alpha_p \omega_p$, we write the spectral density for $\omega \gg \omega_p$ as $J_2(\omega) = E_p \omega_p/\omega$.  (This is quite different to the spectral density for $\omega \ll \omega_p$, which is $J(\omega) = E_p \omega/\omega_p$.)

The second term in \eqref{eq:taylor} is somewhat more difficult to evaluate. We will again Taylor expand in $1/(\tau_p\omega)$. We note, however, that by extracting a factor of $1/a^3$ from both terms of \eqref{eq:taylor} the second term is proportional to $(a/b)^3$, which we already assumed is ``small'' in the first Taylor expansion.   Therefore, when we again Taylor expand around $\omega_p/\omega$ we need only work to zeroth order so that our total final approximation for $J_4(\omega)$ is to first order in two small 
expansion variables.  This yields,
\begin{equation}\label{eq:Js}
J_s(\omega) \approx \frac{9\epsilon_{p,i}}{(2\epsilon_{p,i}+1)^2}
 J_3(\omega)
\end{equation}
where $J_3(\omega)$ is the spectral density for Model 3, with $\epsilon_p \equiv \epsilon_{p,i}$, i.e., the chromophore inside a constant dielectric (high frequency) protein with no cavity, surrounded by the solvent.  The solvent contribution is therefore ohmic, with
a dimensionless coupling constant, $\alpha_s$ given by (\ref{alphas}).


Therefore, equating the two spectral densities suggests a crossover between solvent and protein dominance at frequency $\omega_{co}$,
\begin{equation}\label{eq:cross over}
\frac{\omega_{co}}{\omega_{pe}} = \sqrt{\frac{\alpha_p}{\alpha_{s}}}
\end{equation}
This ratio is always much larger than one for typical values of dielectric constants and relaxation times.

  For frequencies above this limit, provided we are in the regime $b\gg a$, we would expect that the protein dynamics are irrelevant for the system, and the dynamics of the chromophore is
``slaved''
 to the solvent fluctuations.
 Similar effects have been observed in enzyme kinetics\cite{Fenimore02PNAS}. At low frequencies ($\omega \ll \omega_{co}$), the protein dynamics dominate and the details of the solvent are mostly irrelevant. 
 Hence, we expect that even when the chromophore is ``shielded'' from the solvent by the protein that the short time ($\sim 1$ psec) dynamics can still be dominated by the solvent.  This raises questions about the recent assignment of the observed ultra-fast solvation to protein dynamics.\cite{Homoelle98JPCB,Riter96JPCB,Kennis02JPCB} 

\subsection{Model 5 - Bound water}
Our goal is to obtain analytic criteria which tell us when the bound water is relevant. If the dielectric contribution of the bound water dominates over that of the protein, we can use Model 5 to describe the system. Instead of the chromophore pocket being treated as the cavity, now the entire protein is treated as a cavity of radius $b$ with frequency-independent dielectric $\epsilon_p$. This is surrounded by a shell of bound water with radius $c$ (so the shell has width $c-b$) and dielectric $\epsilon_b(\omega)$ (with the subscript representing the bound water). We expect that the layer of bound water (typically \cite{Jordanides99JPCB} about 4.5\AA) will be thin compared to the rest of the protein
($b \sim 20$\AA), and so we are interested in the limit $b \approx c$. 

We simplify the spectral density \eqref{eq:J(omega) model 5} in the same way as Section \ref{sec:Model 4}.  The full details may be found in  the Appendix.
Taylor expanding in $\frac{c-b}{b}$ yields
\begin{align}
J_5(\omega) = J_3(\omega) +  J_{\mbox{bw}}(\omega)
\end{align}
The first term represents the spectral density of a chromophore inside a cavity of radius $b$ with dielectric constant $\epsilon_p$ surrounded by a bulk solvent, and so is the spectral density in the absence of the bound water, as described by Model 3.
 The second term, is proportional
to the ratio $\frac{c-b}{b}$, a measure of the thickness of the bound water,
 and can be identified with the contribution to the spectral density
 of the bound water:
\begin{equation}
 J_{\mbox{bw}}(\omega)
 = \frac{2}{\epsilon_b} 
\left( \frac{2\epsilon_s+\epsilon_b}{2\epsilon_s+\epsilon_p}\right)^2 
\left(\frac{\epsilon_b-\epsilon_s}{2\epsilon_s+\epsilon_b}\right) \frac{(c-b)}{b}
\end{equation}

 Then, the bound water term can be expressed as
\begin{align}
J_{\mbox{\tiny bw}}(\omega)
 &\approx \frac{1}{(2\epsilon_s + \epsilon_p )^2 } \Im \left[ \left( 1+\frac{2\epsilon_{s,s}}{\epsilon_b}\right)(\epsilon_b -\epsilon_{s,s})\right] \\
&= \frac{1}{(2\epsilon_s + \epsilon_p )^2 } \left(1+\frac{2\epsilon_{s,s}^2}{|\epsilon_b(\omega)|^2}\right) \Im \epsilon_b(\omega)
\end{align}
Using a Debye form for the bound water spectral density, gives
Im$\left[\epsilon_b(\omega)\right] =
 (\epsilon_{b,s}-\epsilon_{b,i}) \frac{\omega \tau_b}{1+\omega^2\tau_b^2}$.
 However, we must also include
the $|\epsilon_b(\omega)|^2$ contribution to the frequency dependence. 
If we consider frequencies much less than the bulk solvent relaxation time $1/\tau_s$, then we again find an ohmic spectral density for the bound water
contribution, with dimensionless coupling given by
Equation (\ref{alphab}).

In comparison with the solvent contribution,
\begin{equation}
\frac{\alpha_b}{\alpha_{s}} \approx \left(\frac{c-b}{b} \right)
 \frac{\tau_b}{\tau_s} 
\frac{\epsilon_{b,s} -\epsilon_{b,i}}{\epsilon_{s,s}-\epsilon_{s,i}} 
\frac{(\epsilon_{b,s}^2 +2\epsilon_{s,s}^2)(2\epsilon_{s,i} + \epsilon_p)}{\epsilon_{b,s}^2(2\epsilon_{s,s}+\epsilon_p)}
\end{equation}
We would typically expect $\epsilon_{s,s} \gg \epsilon_{s,i} \;, \; \epsilon_{b,s} \gg \epsilon_{b,i}$, and $\epsilon_{s,s} \gg \epsilon_{b,s}$
 and the protein static dielectric
constant to be small compared to any static frequency,
 but perhaps comparable to the high frequency values of the
dielectric constants of bulk or bound water.
 Therefore,
\begin{align}
\frac{\epsilon_{b,s} -\epsilon_{b,i}}{\epsilon_{s,s}-\epsilon_{s,i}} \frac{(\epsilon_{b,s}^2 +2\epsilon_{s,s}^2)(2\epsilon_{s,i} + \epsilon_p)}{\epsilon_{b,s}^2(2\epsilon_{s,s}+\epsilon_p)}
\sim \frac{\epsilon_{b,s}}{\epsilon_{s,s}} \frac{4 \epsilon_{s,s}^2\epsilon_{s,i} }{2 \epsilon_{b,s}^2 \epsilon_{s,s}} \sim
\frac{\epsilon_{b,i}}{\epsilon_{b,s}}
\end{align}
which we expect to be of order one.
 Therefore, we would usually expect
\begin{equation}
\frac{\alpha_b}{\alpha_{pe}} \sim \frac{c-b}{b} \frac{\tau_b}{\tau_{pe}}
\end{equation}
Hence,
 if $\tau_b \gg \tau_{s}$, as is
observed \cite{Bagchi03ARPC,Yoshiba04B}, then we would expect the bound water to be the dominant effect. Further, since the heights of the peaks are approximately given by their reorganisation energy ($J(\omega_c) = \alpha\omega_c \sim E_R$) we find
\begin{equation}
\frac{E_b}{E_{s}} \sim \frac{c-b}{b}
\end{equation}
where $E_b, E_{s}$ are the reorganisation energies of the bound water and solvent respectively. 

The cross-over frequency between the bound water and bulk water contributions 
dominating the spectral density
can be estimated
by the condition $J_s(\omega_{co}) = J_b(\omega_{co})$.
 Assuming that the bulk and bound water timescales are sufficiently separated that at the cross-over point $J_b(\omega$) is in the decaying tail and $J_e(\omega)$ is in the linear region, then the cross-over frequency is given by
\begin{equation}
\frac{\omega_{co}}{\omega_b} = \sqrt{\frac{\alpha_{s}}{\alpha_b}}
\end{equation}

\section{Spectral densities determined
 from ultra-fast optical spectroscopy }
\label{sec:optical spectroscopy}

The spectral function $J(\omega)$  associated with optical transitions in chromophores can be extracted from ultra-fast laser
 spectroscopy \cite{Fleming96,Wiersma,Pal04CR}.
Two widely used techniques for doing this are the dynamical Stokes shift
and three-photon echo spectroscopy.

{\it Dynamical Stokes shift.}
The time dependence of the Stokes shift in the fluorescence spectrum, where $\nu(t)$ is the maximum (or the first frequency moment) of the fluorescence spectrum at time $t$, can be normalised as
\begin{equation}\label{dynamicstokes0}
C(t) = \frac{\nu(t)-\nu(\infty)}{\nu(0)-\nu(\infty)}
\end{equation}
such that $C(0)=1$, and $C(\infty)=0$ when the fluorescence maxima has reached its equilibrium value.
Using Equation (\ref{eq:nut})
this is related to the spectral density by
\begin{equation}
\label{dynamicstokes}
C(t) = \frac{\hbar}{E\sub{R}} \int_0^\infty 
 d\omega
\frac{J(\omega)}{\omega} \cos(\omega t)
\end{equation}
where $E\sub{R}$ is the total reorganisation energy given in eq. \eqref{eq:reorg}, which also equals the total Stokes shift associated
with solvation.

 The function $C(t)$ is sometimes referred to as the
{\it hydration correlation function}
 and experimental results are often fitted to several decaying exponentials,
\begin{equation}\label{expfit}
C(t) = 
 A_1 \exp(-t/\tau_1) + A_2 \exp(-t/\tau_2)  + A_3 \exp(-t/\tau_3) + \ldots
\end{equation}
where $A_1+A_2+\ldots = 1$.  
From (\ref{dynamicstokes}), this corresponds
 to a spectral density of the form
\begin{equation}
J(\omega) = 
 \frac{\alpha_1\omega}{1+(\omega\tau_1)^2} + \frac{\alpha_2 \omega}{1+(\omega\tau_2)^2} + \ldots
\end{equation}
The dimensionless couplings $\alpha_j$ ($j=1,2,\ldots$) are related
 to the total reorganisation energy by
\begin{equation}
 \alpha_j = \frac{2 E\sub{R} A_j \tau_j}{\pi \hbar}
  \simeq 0.25 A_j \frac{E\sub{R}}{cm^{-1}}
  \frac{\tau_j}{psec}.
\label{alphaj}
\end{equation}

 Table \ref{tab:exp data}
 gives values of the fitting parameters $(E\sub{R},A_j , \tau_j)$
 determined by fast laser spectroscopy for a range of chromophores and different environments, both protein and solvent.
We do not claim the list is exhaustive of all the published values,
but is meant to be indicative (For example, see also \cite{Jimenez94N,Riter96JPCB,Homoelle98JPCB,Changenet-Barret00JPCB,Jimenez03PNAS,Jimenez04PNAS,Cho05JCP} )
 We note the following general features.

 (i) The Stokes shift varies significantly between different environments, both solvent and protein. Generally,
the presence of the protein reduces the total Stokes shift
and the relative contribution of the ultra-fast component,
which can be assigned to the solvent.
The less exposed the chromophore is to the solvent
the smaller is solvent contribution to the spectral density.
This is also seen in measurements of the dynamic Stokes shift
for a chromophore 
placed at three different sites in the B1 domain of protein G.
(See Fig. 3C of Ref. \onlinecite{Cohen02S}).
Denaturing the protein tends to expose the chromophore to more
solvent increase the total Stokes shift and increase the
relative contribution of the ultra-fast component.

 (ii) The different decay times observed for a particular system can vary by as many as four
orders of magnitude, ranging from 10's fsec to a nsec.

 (iii) The relative contributions of the ultra-fast (100's fsec)
 and slow (10's psec) response are often of the same order of magnitude,
consistent with equation (\ref{eq:compare}).

(iv) Even when the chromophores are inside
the protein, the coupling of the chromophore to
the solvent is large.
For example, Prodan is in a hydrophobic pocket of HSA,
well away from the surface, and yet $\alpha_s \sim 50$.
Even for the ``buried'' chromophores 
 (Leu$^7$ and Phe$^{30}$) in GB1,\cite{Cohen02S} the
solvent contribution is $A_s E_R \sim 100$ cm$^{-1}$,
$\tau_s \sim 5 $ psec, and so $\alpha_s \sim 100$.
There are several proteins for which a very slow ($\sim 10$'s nsec)
dynamic Stokes shift has been observed and has been assigned
to dielectric relaxation of the protein 
itself\cite{Pierce92JPC,Pal01JPCB}. 

(v) The values of the dimensionless couplings $\alpha_j$
that we obtain from \eqref{alphaj} are comparable to the rough
estimates we made in Section \ref{overview}.

\begin{table}[p]
\small
\caption{Solvation relaxation times for various chromophores
in different protein and solvent environments. The 
values of relaxation times and their relative weights are 
determined by fitting
 the time dependence of the dynamic Stokes shift (\ref{dynamicstokes0})
to the functional form (\ref{expfit}). 
$E_R$ is the reorganisation energy, given by  
(\ref{eq:reorg}), and equals the total Stokes shift.
Note there is some variation in estimates of the reorganisation
energy depending on whether one estimates it from the maxima
in the absorption and emission spectra or from the
first frequency moment of the spectra \cite{Jordanides99JPCB}.
 It should be noted that the time resolution is
different in the various experiments. Some did not have 
access to femtosecond time scales and so we have left the
relevant columns blank.
SC is Subtilisin Carlsberg.
HSA is Human serum albumin.
SNase-WT is Staphylococcus nuclease in the wild type.
  SNase-K110A is a specific mutant of Staphylococcus nuclease.
HSA is in its native folded form in the buffer but denatures
in concentrations of Gdn.HCl (guanidine hydrogen chloride)
greater than about 5M and at a pH above 7.
 Trp is the amino acid Tryptophan,
Acrylodan is 6-acryloyl-2-(dimethylamino)naphthalene,  Prodan 
is 6-propionyl-2-(dimethylamino)naphthalene,
AP is amino-pyridine,
DCM is 
 4-(dicyanomethylene)-2-methyl-6-(p-dimethylaminostyryl) 4H-pyran,
 MPTS is 8-methoxypyrene-1,3,6-trisulfonate,
 and bis-ANS is 1,1Õ-bis(4-anilino)naphthalene-5,5' disulfonic acid.}
\begin{center}
{
\begin{tabular}{|c|c|c|c|c|c|c|c|c|}
\hline
Chromophore & Protein  & Solvent &  Ref.  & $E_R$ \mbox{(cm$^{-1}$)}  & 
$A_1,\tau_1$ & 
$A_2,\tau_2$ & 
$A_3,\tau_3$  \\
\hline
Trp& none & water & \cite{Zhong02PNAS} &  &   0.65, 160 fsec  & 0.35, 1.1 psec & \\ 
Trp& none & water & \cite{Lu04CPL} & 2193 &   0.55, 340 fsec 
 & 0.45, 1.6 psec &  \\
Trp& SC& buffer
 & \cite{Pal02PNASa} & 1440 &   0.6, 800 fsec  & 0.4, 38 psec & \\
Trp & Monellin & buffer & \cite{Peon02PNAS} & 960  & 0.46, 1.3 psec & 0.54, 16 psec & \\
Trp & SNase-WT & buffer & \cite{Qiu06PNAS} & 850 & 0.46, 5 psec &
 0.54, 153 psec & \\
Trp & SNase-K110A & buffer & \cite{Qiu06PNAS} & 876 &
 0.77, 3 psec & 0.23, 96 psec & \\
Trp & HSA & water, pH 7 & \cite{Qiu06JPCB} & 1156 &
 0.39, 5 psec & 0.61, 133 psec & \\
Trp & HSA & water, pH 9 & \cite{Qiu06JPCB} & 1015 &
 0.3, 1.6 psec & 0.7, 46 psec & \\
Dansyl & SC& water & \cite{Pal02PNASa} & 1180 & 0.94, 1.5 psec  & 0.06, 40 psec  &\\
DCM & HSA & Tris buffer & \cite{Pal01JPCB} &  515 & & 0.25, 600 psec  & 0.75, 10 nsec \\
Prodan & none & buffer & \cite{Kamal04PNAS} & 2313  &  0.47, 130 fsec & 0.53, 770 fsec &  \\
Prodan & HSA & buffer & \cite{Kamal04PNAS} & 916   &  0.19, 780 fsec & 0.56, 2.6 psec &  0.25, 32 psec \\
Acrylodan & HSA & buffer & \cite{Kamal04PNAS} & 1680  & 0.23, 710 fsec & 0.41, 3.7 psec & 0.36, 57 psec \\
Acrylodan & HSA & 0.2M Gdn.HCl & \cite{Kamal04PNAS} &   & 0.16, 280 fsec & 0.36, 5.4 psec & 0.48, 61 psec \\
Acrylodan & HSA & 0.6M Gdn.HCl & \cite{Kamal04PNAS} &  & 0.2, 120 fsec & 0.55, 2 psec & 0.25, 
13.5 psec \\
MPTS & none & buffer  & \cite{Jimenez02JPCB} & 2097  & 0.8, 20 fsec  & 0.2,  340 fsec    &  \\ 
MPTS & Ab6C8 & buffer & \cite{Jimenez02JPCB} & 1910  & 0.85, 33 fsec  & 0.1,  2 psec    &  0.05, 67 psec\\ 
bis-ANS & GlnRS (native) & water  & \cite{Sen03JPCB} &  750  && 0.45, 170 psec  & 0.55,  2.4 nsec     \\ 
bis-ANS & GlnRS (molten) & urea  & \cite{Sen03JPCB} &  500  && 
0.63, 60 psec  & 0.37,  0.96 nsec     \\ 
4-AP & GlnRS (native) & water  & \cite{Sen03JPCB} &  1330 &&
0.85, 40 psec  & 0.15,  580  psec      \\ 
4-AP & GlnRS (molten) & urea  & \cite{Sen03JPCB} &  700 &&
0.77, 50 psec  & 0.23,  0.9 nsec      \\ 
Zn-porphyrin & Cytochrome-c & water & \cite{Lampa04JPCB} &  170 & &
0.4, 250 psec  & 0.6,  1.5 nsec      \\ 
\hline
\end{tabular}
} 
\end{center}
\label{tab:exp data}
\end{table}


{\it Three pulse photon echo spectroscopy.}
 This technique is analogous to stimulated spin echo measurements
used in nuclear magnetic resonance to extract the phase relaxation time $T_2$.
For ``long'' times it can be shown\cite{Wiersma} that
the time-dependent echo peak shift $S(t)$, where $t$ is the
time delay between the second and third pulse,
is related to the correlation function $C(t)$, given in
(\ref{dynamicstokes}), by
\begin{equation}
\label{eq:echo}
S(t) = {\tau_g  \over \sqrt{\pi} } C(t)
\end{equation}
where $\tau_g$ is the decoherence
timescale  given in eq. \eqref{eq:gaussian}, and associated with the
``collapse of the wave function.''

 The solvation dynamics of the fluorescein dye eosin bound to lysozyme in an aqueous solution was studied and compared to that
 for eosin in water without
the protein\cite{Jordanides99JPCB}.
 For both systems, ultra-fast solvation relaxation occurs in about 10 fsec and is assigned to bulk water. However, for the lysozyme-eosin complex
a slower relaxation also occurred on the scale of 100 psec.
 This was assigned as predominantly due to water bound to the protein,
 mostly in the first hydration shell. This can be compared with dielectric dispersion measurements \cite{Harvey72JPC} which suggest that there are two solvation relaxation times
of 4 psec and 270 psec. A molecular dynamics simulation of
 lysozyme in an explicit solvent environment of 5345 water molecules found a single solvation relaxation time of 100 psec
 \cite{Smith93JPC}.
Jordanides {\it et al.}\cite{Jordanides99JPCB} used the dynamic
 dielectric continuum model of
 Song and Chandler \cite{Song98JCP} to extract the 
spectral density, based on four different dielectric models.
 The full time dependence of the solvation was best described by a model
 which included the frequency dependence of dielectric constant of both
the lysozyme  and the water bound at the protein surface.
 These models for the lysozyme complex can be compared
to our models if some simplifying assumptions are made. 
In particular, we need to
treat the lysozyme protein as spherical
with the eosin complex at its centre. Models I and II in
Ref. \cite{Jordanides99JPCB}
then correspond to our Models 3
and 5, respectively.
Models III and IV are approximately our Model 4, with the appropriate choice for the protein dielectric constant of the protein.

\section{Comparison with spectral densities determined
 from molecular dynamics simulations} 
\label{sec:simulations}

For several specific proteins
molecular dynamic simulations have been used to
determine several quantities relevant to this
work: the static dielectric constant of the protein,
the frequency dependent dielectric constant,
solvation dynamics, or
the spectral density associated with an
optical transition in a chromophore or
an electron transfer.
\cite{Smith93JPC,Pitera01,Hofinger01,Loffler97,Xu92,Kosztin06,Boresch00JPCB,Warshel01QRB,Miyashita00JPCB,Rudas06JCP,Nilsson05PNAS}
We hope that our work will stimulate further
simulations of the spectral density for specific chromophores 
and proteins in an aqueous environment.
To determine it one needs to
calculate time correlations of the (reaction field)
electric field at the location of the chromophore
within the protein.
Equivalently,
the spectral density can be related to the fluctuations in the energy difference between the ground and excited states of the system \cite{Xu92}.
We now briefly review some of the results on specific proteins
that are relevant to this paper.

\subsection{Tryptophan in monellin and water}
Molecular dynamics was used to calculate the time correlation function
 $C(t)$ for trajectories of a  few nanoseconds \cite{Nilsson05PNAS}. 
 For free Trp in bulk water $C(t)$ was fit to a bi-exponential decay function with $A_1 = 0.86 \pm 0.04, \tau_1 = 70 \pm 10$ fs,
 and $A_2 = 0.14 \pm 0.04, \tau_2 = 0.7 \pm 0.2$ ps. 
 For Trp-3 in the protein monellin $C(t)$ was
 fit to a tri-exponential form with $A_1 = 0.66 \pm 0.02, \tau_1 = 70\pm 10$ fs; $A_2 = 0.22 \pm 0.02, \tau_2 = 1.0 \pm 0.1 $ps; $A_3 = 0.12 \pm 0.01,\tau_3 = 23\pm 2$ psec.
The total reorganisation energy was calculated
as $E_R = 3200$ cm$^{-1}$.
  The two faster decays were assigned to the bulk water and the slowest component ($\tau_3 =  23 \pm 2$ ps) was assigned to protein dynamics including the motion of the chromophore within the protein.
  This assignment is consistent with the interpretation of NMR measurements but is different to that given in the associated experimental measurements \cite{Peon02PNAS} of the time dependent Stokes shift.
  The latter assigned the slower time scale ($\sim 20$ psec)
 to the dynamic exchange between water bound at the protein surface (the first hydration shell) with bulk water.

\subsection{Protein GB1 in water}

The dynamic Stokes shift of a chromophore at
the site of 
several different amino acid residues within 
the B1 domain of the protein G was measured.\cite{Cohen02S}
The residues were replaced with an Aladin chromophore
at sites that were ``buried''
 (Leu$^7$ and Phe$^{30}$),
partially exposed (Trp$^{43}$), and exposed(Ala$^{24}$),
to the solvent.
The more exposed the site the larger the
dynamic Stokes shift and the faster the relaxation.
Motivated by these experiments,
Golosov and Karplus performed molecular
dynamic simulations for this 56 residue protein
in a solvent of 6205 water molecules.\cite{Golosov07JPCB}
They calculated the time-dependent correlation
function for the electrostatic interaction energy
of the site residue with the rest of the system.
This quantity should scale with the energy gap correlation
function.
For eleven different sites the solvent coverage (defined as
the ratio of the surface area of the residue that
is accessible to the solvent to the surface area of the
isolated residue) ranged from 5 to 45 per cent. 
The hydration correlation function  was found
to vary significantly between sites, but
all contained components that could
be assigned to ultra-fast decay (on the 100 fsec 
and 1 psec timescales,
due to the surrounding water) and much
slower relaxation (on the 100's psec timescale)
that could be assigned to 
coupled hydration and protein conformational dynamics.
However, there was no simple correlation between
the slow relaxation time scale and the extent of the
exposure of the site to the solvent,
contrary to the correlation found by others.\cite{Bagchi05JACS}

\subsection{Frequency dependent dielectric
properties of an HIV1 zinc finger peptide in water
}\label{sec:HIV1}

This peptide consisted of  18 amino
acid residues and was simulated in a periodic
box containing 2872 water molecules.\cite{Loffler97}
 It was simulated for 13.1 nsec and exhibited a clear 
separation of time scales associated with dielectric relaxation
of the different parts of the system.
The water had a dielectric relaxation time
of 7 psec, comparable to that for bulk water.
 Dielectric relaxation of the protein was
dominated by a time scale of 4.3 nsec comparable
to that found in simulations of other proteins and
comparable to the time scale for rotation of the whole protein.
The static dielectric constant of the peptide was estimated to
be 15.

\subsection{Frequency dependent dielectric
properties of ubiquitin   in                  water
}\label{sec:ubiquitin}

Ubiquitin is a small globular protein composed of
76 amino acids. It was simulated in a cubic box containing
5523 water molecules for runs of
5 nsec duration.\cite{Boresch00JPCB}
Time dependent correlation functions (which are the
fourier transform of the frequency dependent
dielectric constant) could be fit to 
sums of two decaying  exponentials with different
weights and relaxation times.
For the dielectric relaxation the three dominant timescales
observed were 7 psec, 2.6 nsec, and 1.9 nsec.
These were associated with the bulk water, with rotation of the
whole protein, and the bound water and side chains at the
protein surface, respectively.
Recently the same group extended the simulations to
20 nsec and also calculated the frequency dependent
dielectric constant of solutions of
the proteins apo-calbindin $D_{9K}$ and the
C-terminal domain of phospholipase C-$\gamma 1$.\cite{Rudas06JCP}

\section{Conclusions}

The focus of this paper has been
on the coupling of optical transitions in biological chromophores
to their environment. However,
  the approach and results presented here can be readily
adapted to other transitions involving two quantum states
which differ in the value of their electric dipole moment.
Examples include intersystem crossing, non-radiative
decay via a conical intersection, electron transfer
and proton transfer.

We hope our work will stimulate
more work considering  the following general claims,
which this paper has elucidated.

(i) A valuable                      approach
to  modelling quantum dynamics in specific
biomolecular systems may be in terms of
``minimal'' models such as the spin-boson model where
the system parameters and spectral density are extracted from
experiment and/or quantum chemistry and molecular dynamics.

(ii) Even when the active site of a biomolecule is shielded from 
bulk water, the latter     can still have a significant effect
on the quantum dynamics of the active site, especially if the
time scale of interest is comparable to the solvation time
scale associated with the bulk water.
This can lead to solvent fluctuations dominating protein
dynamics and function \cite{Fenimore02PNAS}.

(iii) The environment of the active site can be divided into three
distinct components: the surrounding protein, water at the 
protein surface, and bulk water. The times scales
associated with the dielectric relaxation of
each component usually differs by several orders of magnitude
and so each makes a unique contribution 
the coupling of the quantum dynamics of the active site to the
environment. Furthermore, the relative importance
of each component depends on how the time (or energy)
scale of the quantum dynamics,
compares to time scale of the solvation associated
with each of the components of the environment.
Table 1 compares the associated energy scales.

(iv) The time scales associated with decoherence and the
``collapse of the wave-function'' in these biomolecular systems
are experimentally accessible. Given the high tuneability of
these systems they could potentially be used in
fundamental studies concerning  quantum measurement theory.

\section{Acknowledgements}

This work was supported by the Australian Research Council.
  We thank Paul Burn, Jacques Bothma, Minhaeng Cho, Dan Cox,
Paul Curmi, Paul Davies, Andrew Doherty, Ken Ghiggino,
Noel Hush, Martin Karplus, Alan Mark,
Hugh McKenzie, Paul Meredith, Gerard Milburn, Seth Olsen, Samir Pal,
Ben Powell, Jeff Reimers,
Jenny Riesz, Maximilian Schlosshauer, Greg Scholes,
Thomas Simonson, Rajiv Singh, Jeff Tollaksen, and Dongping Zhong
 for very helpful discussions.



\begin{appendix}
\section{Timescales}

\begin{table}[htdp]
\caption{Timescales for various processes in biomolecules and solutions.  The radiative lifetime of a chromophore is order of magnitudes longer than all other timescales, except perhaps protein dielectric relaxation.  MD refers to results from molecular dynamics simulations. Of particular relevance to this work is the separation of timescales, $\tau_s \ll \tau_b \ll \tau_p$
(compare Fig. \ref{fig:timescales}).}
\begin{center}
\begin{tabular}{|c|c|c|}
\hline
Process & Timescale &  Ref.\\
\hline
Radiative lifetime & 10 ns &    \cite{vanHolde98} \\
Internal conversion & 10fs & \cite{vanHolde98} \\
Bulk water dielectric relaxation & 8 ps  & \cite{Afsar78} \\
Protein dielectric relaxation (MD), $\tau_{D,p}$ & 1-10 ns  & \cite{Loffler97,Boresch00JPCB} \\
Ultrafast solvation in water & 10's fs  &  \cite{Lang99JCP}\\
Fast solvation in water, $\tau_s$ & 100's fs & \cite{Lang99JCP}  \\
Solvation due to bound water, $\tau_b$ &  5-50 ps & \cite{Peon02PNAS} \\
Solvation due to protein, $\tau_p$& 1-10 ns & \cite{Sen03JPCB} \\
Covalent bond vibrations & 10-100 fs & \cite{vanHolde98}\\
Elastic vibrations of globular regions & 1-10 ps & \cite{vanHolde98} \\
Rotation of surface sidechains & 10-100 ps & \cite{vanHolde98} \\
Reorientation of whole protein & 4-15 ns  & \cite{Boresch00JPCB}\\
\hline
\end{tabular}
\end{center}
\label{tab:timescales}
\end{table}

\section{Solution for the reaction field}

\begin{figure}[htbp]
\begin{center}
\includegraphics[width=3.5in]{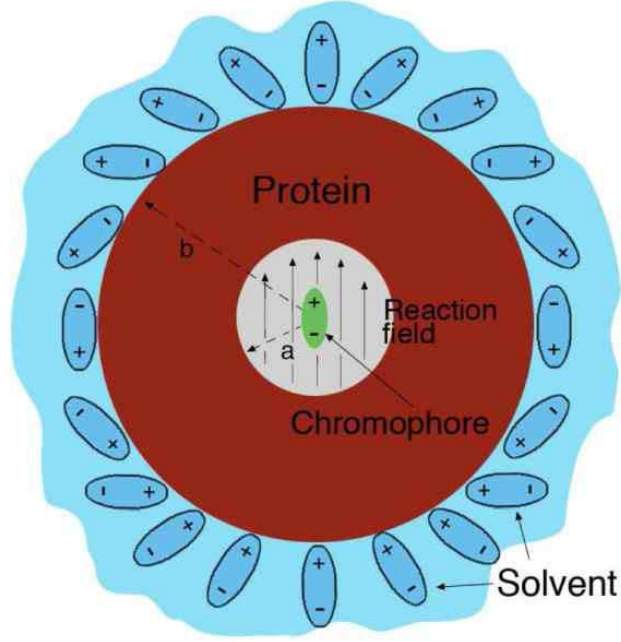}
\caption{Model 4 for the interaction between a chromophore and its environment.  The chromophore is treated as a point dipole sitting in a cavity of radius $a$ in the centre of a spherical, uniform protein which is treated as a homogeneous dielectric medium of radius $b$.  The protein-pigment complex is surrounded by a solvent, typically water, which is again treated as a homogeneous dielectric medium though actual molecules are shown for clarity of explanation.  The chromophore's dipole moment polarises its environment, which in turn produces an electric field, the "reaction field", which interacts with the chromophore.  Fluctuations in the environment will translate to fluctuations in the chromophore's energy.}
\label{fig:model 4}
\end{center}
\end{figure}

A change in the dipole moment of the chromophore leads
to a re-organisation of the environment, which produces a
reaction field acting back on the dipole. This is shown 
schematically in Figure \ref{fig:model 4}.

The reaction field for these models can be obtained by a generalisation of the techniques in ref. \cite{Bottcher73} as follows: The electric potential, $\phi(x,y,z)$, satisfies Poisson's equation $\grad^2 \phi=  - \rho/\epsilon$, where $\epsilon$ is the local dielectric constant of the medium and $\rho$ is the local charge density. Away from the point dipole and surface boundaries, $\rho=0$ and we must solve Laplace's equation $\grad^2 \phi=0$.  At the dielectric boundaries (and in general), $\phi$ must be continuous and because there are no free charges, $\vec{D}=\epsilon \vec{E} = -\epsilon \grad \phi$ is also continuous across the boundaries.

Although the protein is spherically symmetric, because of the point-dipole, the system has only cylindrical symmetry. If the spherically symmetric electric potential in each concentric dielectric shell is given by $\phi_1(r), \phi_2(r), \ldots$, we can expand $\phi_i$ in terms of spherical harmonics \cite{Bottcher73}:
$$\phi_i= \sum_{n=0}^\infty \left( A_{i,n} r^n + \frac{B_{i,n}}{r^{n+1}} \right) P_n(\cos \theta)$$
We consider explicitly the case where we have a cavity surrounded by a single dielectric shell inside a bulk solvent. The central cavity has radius $a$, dielectric $\epsilon_c$ and potential $\phi_c(r,\omega)$, the shell has total radius $b$ (thickness $b-a$),
dielectric $\epsilon_p$ and potential $\phi_p(r,\theta)$ , and the bulk environment is described by dielectric $\epsilon_e$ and potential $\phi_e(r,\omega)$.

We can then apply the boundary conditions:
\begin{subequations}
\begin{align}
\phi_e(r\rightarrow\infty)  & \rightarrow 0 \\
\phi_\mu  &= \frac{\mu}{r^2}\cos\theta \\
(\phi_p)_{r=b} &= (\phi_e)_{r=b} \\
(\phi_c)_{r=a} &= (\phi_p)_{r=a} \\
\epsilon_c \left( \frac{\partial \phi_c}{\partial r} \right)_{r=a} &= \epsilon_p \left( \frac{\partial \phi_p}{\partial r} \right)_{r=a} \\
\epsilon_p \left( \frac{\partial \phi_p}{\partial r} \right)_{r=b}&= \epsilon_e \left( \frac{\partial \phi_e}{\partial r} \right)_{r=b}
\end{align}
\end{subequations}
The first condition is that the potential must go to zero at infinity.  This means that all coefficients with positive powers of $r$ must vanish, i.e., $A_{e,n}=0$ for all $n$.

The second condition is the field from a point dipole. As this is the only free charge in the cavity, this is the only source term (inverse power of $r$) that will contribute to the potential $\phi(r, \theta)$. Since $P_1(\cos \theta) = \cos \theta$, only the $n = 1$ term is involved. Therefore, $B_{c,n=1} = \mu$ and $B_{c,n\neq 1} = 0$. (Nothing is said about $A_{c,n}$).

The final terms describe the continuity of the potential and its derivative over the boundary. The first condition
gives
\begin{equation}
\sum_{n=0}^\infty \left( A_{p,n} b^n + \frac{B_{p,n}}{b^{n+1}}\right) P_n (\cos\theta) = \sum_{n=0}^\infty \frac{B_{e,n}}{b^{n+1}} P_n (\cos\theta)
\end{equation}
Because the spherical harmonics $P_n$ are orthogonal, we can consider each term of this sum as being equal, so
\begin{equation}
A_{p,n} b^n + \frac{B_{p,n}}{b^{n+1}} = \frac{B_{e,n}}{b^{n+1}}
\end{equation}
In a similar way, the remaining boundary conditions can be applied to produce a set of linear equations on the $A_{i,n}$ and and $B_{i,n}$. We have six boundary conditions and six variables (each, of course, a function of $n$) and so we are able to solve for all parameters. However, we are only interested in the field inside the cavity, and in particular the unknown part $A_{c,n}$. We find that all the $A_{c,n}$ are zero except for $n = 1$. Thus, the potential due to the surface charges is given by $\phi_{c,\mbox{\tiny surf}} = - \chi\mu r \cos\theta = -\chi \mu \op{z}$ where we find
\begin{equation}\label{eq:chi omega}
\chi(\omega) = \frac{2}{a^3}
\frac{ (\epsilon_p+2\epsilon_c)(\epsilon_e-\epsilon_p) a^3 + (\epsilon_p-\epsilon_c)(2\epsilon_e + \epsilon_p) b^3}{ 2(\epsilon_p-\epsilon_c)(\epsilon_e-\epsilon_p) a^3 + (2\epsilon_p+\epsilon_c)(2\epsilon_e+\epsilon_p) b^3}
\end{equation}
The actual electric field in the cavity due to the surface charges but not the dipole itself -- the reaction field -- is then
 $\vec{R}= R \op{z} = -\grad\phi_{e,\mbox{surf}}(x, y, z) = \chi \mu \op{z}$,
 which will be a constant throughout the cavity, 
parallel to the dipole, and proportional to the dipole moment $\mu$. The spectral density describing coupling of 
changes in 
the chromophore state to this environment is related to the zero temperature fluctuations in the reaction field \cite{Gilmore05JPCM}:
\begin{equation}
J(\omega)  = (\Delta\mu)^2 \Re \int dt \: e^{i\omega t} \expec{R(t)R(0)}_{T=0}
\label{J-react}
\end{equation}
This can be shown by writing the reaction field $R(t)$ in terms of its normal modes'  creation and annihilation operators. $\Delta\mu$ is the change in chromophore dipole moment on the transition from the ground to excited states. The fluctuations
in the reaction field $\expec{R(t)R(0)}$ are obtained \cite{Gilmore05JPCM} from the fluctuation dissipation theorem, and are proportional to
the imaginary part of $\chi(\omega)$ in \eqref{eq:chi omega} above, yielding 
\begin{equation}
J(\omega) = 2(\Delta\mu)^2 \Im(\chi(\omega))
\end{equation}
Note the use of zero temperature fluctuations is a mathematical derivation only, and provided the appropriate temperature parameters for the solvent and protein are used, the resulting spectral density is applicable to all temperatures.

\end{appendix}
\end{document}